\documentclass{article}
\usepackage{spconf}
\usepackage{amsmath,amssymb,graphicx,booktabs,multirow}
\usepackage[hidelinks]{hyperref}

\newcommand{\demourl}{https://pnlong.github.io/jazzsamba-demo/}
\newcounter{demofn}

\makeatletter
\def\name#1{\gdef\@name{#1}}
\def\@maketitle{%
  \newpage
  \null
  \vskip 2em
  \begin{center}
    {\large \bf \@title \par}%
    \vskip 1.5em
    {\large
      \begin{tabular}{@{}c@{}}
        \@name\\[1em]
        \@address
      \end{tabular}\par}%
  \end{center}
  \par
  \vskip 1.5em}
\makeatother

\title{JazzSAMBA: A Synchronous and Asynchronous Multi-take Band Audio Dataset of Jazz Standards for Live Music Models}
\name{%
{\em\begin{tabular}{@{}c@{}}
Phillip Long$^{1}$,
Jacob Nguyen$^{3}$,
Jace Hosto$^{3}$,
Gage Hosto$^{3}$,
Jett Takazawa$^{3}$,
Fares Nofal$^{3}$,
Sebastian Stade$^{3}$,\\
Nithya Shikarpur$^{2}$,
Julian McAuley$^{1}$,
Cheng-Zhi Anna Huang$^{2}$,
Stephen Brade$^{2,*}$,
Aleksandra Teng Ma$^{2,*}$\thanks{Corresponding author: \texttt{p1long@ucsd.edu}.
$^{*}$Equal contribution.}%
\end{tabular}}}
\address{%
$^{1}$University of California, San Diego \quad
$^{2}$Massachusetts Institute of Technology \quad
$^{3}$Independent Musician}

\begin{document}

\maketitle

\begin{abstract}
Machine learning has made strong progress on music tasks, both as assistive
tools and as creative partners.
However, most systems train on multitrack corpora that emphasize pop and rock.
Jazz, with improvisation at the core of its practice, still lacks a
well-annotated corpus of clean per-stem combo recordings on standards.
We introduce \textbf{JazzSAMBA} (\textbf{J}azz \textbf{S}ynchronous
and \textbf{A}synchronous \textbf{M}ulti-take \textbf{B}and \textbf{A}udio) to
fill this gap: the first originally recorded jazz-combo multitrack dataset of
standards with asynchronous (overdubbed) and synchronous (live ensemble)
protocols, preferred and alternate takes chosen by the musicians, and timed
annotations for bars, chords, sections, and soloists.
JazzSAMBA covers 76 standards by eight musicians on drums, bass, piano,
trumpet, and saxophone, with per-stem audio, mixtures, and MIDI.
It can support chart-conditioned accompaniment, combo source separation, and
form-aware music information retrieval.
We demonstrate the dataset on two tasks: a jazz combo source-separation
baseline and a chart-conditioned accompaniment ablation.
The dataset, code, and samples are linked from the project demo
page.\footnote{\url{\demourl}}\setcounter{demofn}{\value{footnote}}
\end{abstract}


\section{Introduction}
\label{sec:intro}

Machine learning has made strong progress on music understanding, live music
generation, and other related music tasks.
However, most systems train on multitrack corpora that emphasize pop and
rock~\cite{musdb18,pereira2023moisesdb,bittner2014medleydb}.
As a result, these datasets usually limit tasks to standard pop and rock
instruments and styles, which usually involve compositions and arrangements
created beforehand.
Jazz, on the other hand, combines pre-composed charts with improvisation, so
the improvisation space is bounded by harmony and structure rather than left
unconstrained.
Prior jazz resources cover audio-aligned harmony, formal structure, solo
transcription, and solo-piano or stem-transcribed
performances~\cite{eremenko2018audio,balke2022jsd,pfleiderer2017inside,row2023jazzvar},
yet they rarely provide originally recorded multi-instrument combo stems of
standard repertoire~\cite{cheston2024jazz}.
However, interactive accompaniment, real-time generation, and source separation
need corpora with clean ensemble stems. Chart-conditioned generation and form-aware
analysis further benefit when those stems are paired with structural annotations
that models can condition on or score against.

We propose JazzSAMBA, the first jazz-standard dataset with both clean per-stem combo
recordings and performance-aligned chord and section labels.
It provides asynchronous (overdubbed) and synchronous (live ensemble)
protocols, preferred and alternate takes chosen by the musicians, and timed
bar and soloist annotations.
It comprises 76 standards by eight musicians on drums, bass, piano, and horns.
These assets support chart-conditioned accompaniment, combo-standard
instrument source separation, and form-aware MIR.
We demonstrate two uses that depend on these assets: a jazz 
combo source-separation baseline, and a chart-conditioned causal accompaniment 
benchmark that ablates timed sections and chords against listen-only conditioning.
The dataset, code, and samples are linked from the project demo
page.\footnotemark[\value{demofn}]

\begin{figure*}[t]
  \centering
  \includegraphics[width=\textwidth]{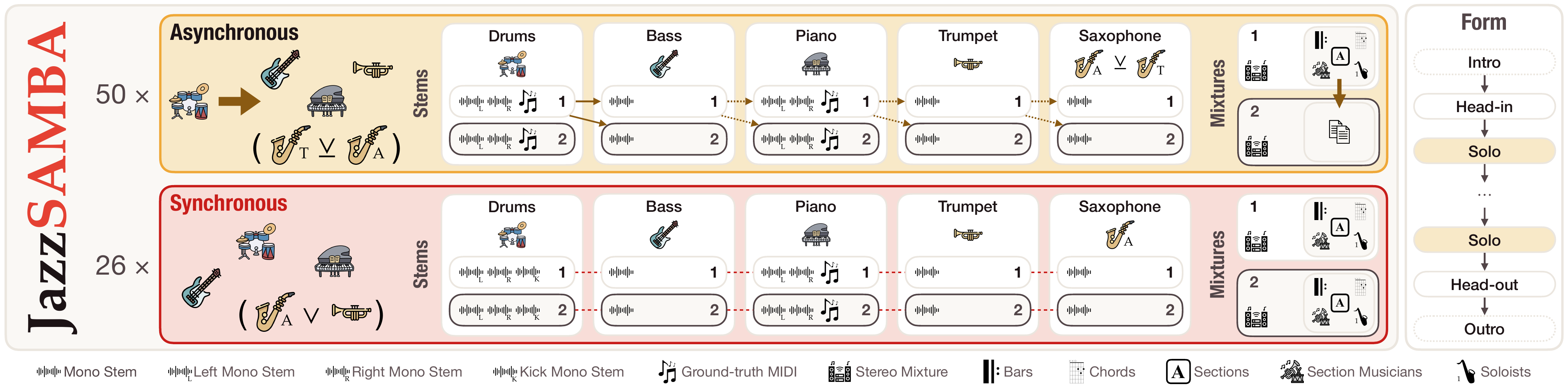}
  \caption{JazzSAMBA overview ($50$ asynchronous / $26$ synchronous songs;
  about $8.1$\,h preferred-take, about $16.1$\,h both tiers).
  Labels 1 and 2 are take tiers.
  Asynchronous takes overdub in sequence (drums first; dotted arrows);
  synchronous takes record the whole ensemble together (dashed lines).
  Preferred mixtures are annotated for bars, chords, sections, and soloists;
  alternate async annotations are copied, while sync takes are labeled
  separately.}
  \label{fig:overview}
\end{figure*}

\section{Related Work}
\label{sec:related}

\vspace{-1.5em}

\begin{table}[h]
  \centering
  \caption{JazzSAMBA vs.\ selected related corpora.}
  \label{tab:comparison}
  \footnotesize
  \setlength{\tabcolsep}{3.2pt}
  \resizebox{\columnwidth}{!}{%
  \begin{tabular}{@{}lccccccc@{}}
    \toprule
    \textbf{Corpus} & \textbf{Stems} & \textbf{Takes} & \textbf{Dual} & \textbf{Chords} & \textbf{Sections} & \textbf{Soloists} & \textbf{MIDI} \\
    \midrule
    MUSDB18~\cite{musdb18} & \checkmark & & & & & & \\
    MoisesDB~\cite{pereira2023moisesdb} & \checkmark & & & & & & \\
    Slakh2100~\cite{manilow2019cutting} & \checkmark & & & & & & \checkmark \\
    Jazz Trio Database~\cite{cheston2024jazz} & \checkmark$^\dagger$ & & & & & & \checkmark$^*$ \\
    Jazz Harmony~\cite{eremenko2018audio} & & & & \checkmark & \checkmark & & \\
    Weimar Jazz Database~\cite{pfleiderer2017inside} & & & & \checkmark & \checkmark & \checkmark & \checkmark$^*$ \\
    Jazz Structure Dataset~\cite{balke2022jsd} & & & & & \checkmark & \checkmark & \\
    RWC Jazz~\cite{balke2026rwc} & & & & & \checkmark & & \checkmark \\
    JAZZVAR~\cite{row2023jazzvar} & & & & \checkmark & & & \checkmark$^*$ \\
    H2H Music Improv~\cite{ma2026h2h} & \checkmark & \checkmark & & & & & \\
    \midrule
    \textbf{JazzSAMBA}
      & \checkmark & \checkmark & \checkmark & \checkmark & \checkmark & \checkmark & \checkmark \\
    \bottomrule
  \end{tabular}}\\[0.35em]
  \raggedright
  Dual: asynchronous and synchronous protocols.
  $^\dagger$Separation-derived stems;
  $^*$MIDI from automatic or monophonic transcription.
\end{table}

\noindent\textbf{Multitrack datasets.}
Studio and synthetic multitracks (MUSDB18, MedleyDB, MoisesDB, Slakh2100)
support separation and transcription outside
jazz~\cite{musdb18,bittner2014medleydb,pereira2023moisesdb,manilow2019cutting}.
JazzSAMBA targets that gap with originally recorded jazz-combo stems of
standards rather than pop or rendered MIDI.

\noindent\textbf{Jazz corpora.}
Table~\ref{tab:comparison} compares related jazz and improvisation audio
corpora along stems, takes, recording protocols, annotations, and MIDI.
Prior jazz resources emphasize harmony and form labels, solo transcription,
or solo piano without originally recorded multi-instrument combo
stems~\cite{eremenko2018audio,balke2022jsd,pfleiderer2017inside,row2023jazzvar,balke2026rwc};
where MIDI is present, it is often automatic or monophonic, and the jazz trio
database supplies stems only via source
separation~\cite{cheston2024jazz}.
H2H Music Improv provides multi-take free-improvisation per-stem audio with
bidirectional intention annotations~\cite{ma2026h2h}, but it is non-idiomatic
and does not constrain the corpus to any specific style.
JazzSAMBA unifies close-mic combo stems, dual protocols and takes, and timed
chart annotations on standards.

\noindent\textbf{Interactive accompaniment.}
Offline models include the Anticipatory Music
Transformer~\cite{thickstun2023anticipatory}; online systems include RL-Duet,
RealChords, StreamGen, RealJam, and
LiveBand~\cite{jiang2020rl,wu2024realchords,wu2025streaming,scarlatos2025realjam,pasini2026liveband},
mostly on symbolic pop/classical or non-jazz audio.
Jazz suits such models: improvisation on shared charts couples freedom with
form models can follow. JazzSAMBA could expand these models to real combo
scenarios with its clean stems and chord and section annotations.

\section{JazzSAMBA Dataset}
\label{sec:jazzsamba}

\begin{figure}[h]
  \centering
  \includegraphics[width=\columnwidth]{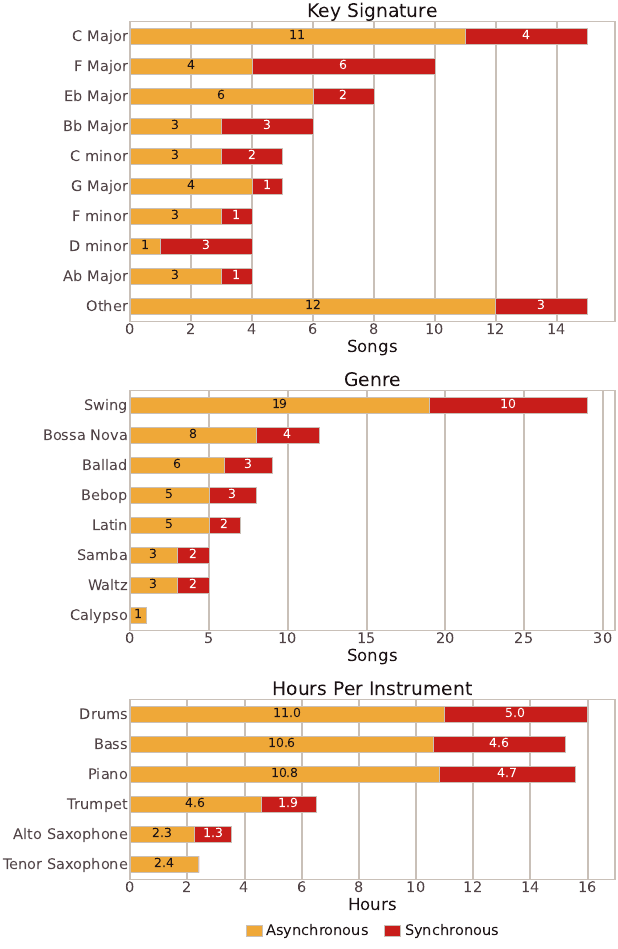}
  \caption{JazzSAMBA corpus statistics by protocol (asynchronous or synchronous):
  key signatures, genres, and active hours per instrument from section
  musician annotations over both take tiers, excluding silence while sitting out.}
  \label{fig:corpus}
\end{figure}

JazzSAMBA uses two recording protocols: instruments tracked one at a time, and
the full band tracked together
(Figure~\ref{fig:overview}; Table~\ref{tab:protocols}).
In the \emph{asynchronous} protocol, musicians record separately: drums first
establish a rhythmic foundation, then later instruments overdub on preferred
earlier takes.
Each instrument records two clean takes; preferred stems form the preferred mixture
and the remaining takes form the alternate tier.
Fixed-tempo overdubs let us annotate the preferred mixture once and copy those
labels onto the alternate tier.
In the \emph{synchronous} protocol, the full band records two ensemble takes,
musicians pick one preferred take, and close-mic bleed makes this the harder
condition; takes are annotated independently.

\begin{table}[t]
  \centering
  \caption{Asynchronous vs.\ synchronous protocol contents.
  Shared marks preferred labels copied to the alternate tier.
  GT denotes ground-truth; remaining MIDI is from MuScriptor
  AMT~\cite{rouard2026muscriptor}.
  For stems with bleed, we also ship a Wiener-style de-bleeded
  version~\cite{kokkinis2012wiener}.
  All audio is $48$\,kHz.}
  \label{tab:protocols}
  \footnotesize
  \setlength{\tabcolsep}{2.4pt}
  \resizebox{\columnwidth}{!}{%
  \begin{tabular}{@{}l|cc|cc|cc@{}}
    \toprule
    \multirow{2}{*}{\textbf{Protocol}}
      & \multicolumn{2}{c|}{\textbf{Stems}}
      & \multicolumn{2}{c|}{\textbf{Annotations}}
      & \multicolumn{2}{c}{\textbf{GT MIDI}} \\
    \cmidrule(lr){2-3} \cmidrule(lr){4-5} \cmidrule(l){6-7}
      & \emph{Clean} & \emph{Bleed}
      & \emph{Timed} & \emph{Shared}
      & \emph{Drums} & \emph{Piano} \\
    \midrule
    Asynchronous
      & \checkmark & & \checkmark & \checkmark & \checkmark & \checkmark \\
    Synchronous
      & & \checkmark & \checkmark & & & \checkmark \\
    \bottomrule
  \end{tabular}}
\end{table}

Both protocols store the same timed annotations on a shared bar grid.
Bars give onset times and measure indices; chords, sections, and soloists use
bar and fraction offsets so harmony and form stay aligned to the performance.
Chord labels come from written lead sheets, not audio estimates:
MusPyExpress~\cite{long2025muspyexpress} parses an internal chart collection,
then musicians time-align them, keeping the written symbol and fields such as
root, triad, seventh, extensions, alterations, and slash bass
(e.g.\ \textit{E-7b5}).
Section labels use jazz form tags (intro, head-in, head-out, solos, outro),
with nested letters when needed (e.g.\ \textit{head-in/A}).
A per-section musician table records who plays; soloist rows give an ordered
solo sequence.

The corpus covers 76 standards by eight musicians on drums, bass, piano,
trumpet, and saxophone (alto or tenor, exclusive per song): 50 asynchronous
and 26 synchronous.
Repertoire spans swing, bossa nova, ballad, bebop, and latin feels, mostly in
$4/4$ with both swung and straight grooves.
Preferred mixtures total about $8.1$\,h ($5.5$\,h asynchronous,
$2.5$\,h synchronous); both tiers reach about $16.1$\,h at $75$--$270$\,BPM.
Figure~\ref{fig:corpus} summarizes keys, genres, and hours per instrument;
listening examples are on the demo page.

\section{Experiments}
\label{sec:experiments}
We demonstrate JazzSAMBA in two settings: source separation and chart-conditioned
accompaniment.
Source separation fine-tunes and evaluates on asynchronous stem-sum mixtures
for drums, bass, piano, and horns.
Chart-conditioned accompaniment trains on both protocols and ablates
oracle-timed chord and section masks against listen-only conditioning on
asynchronous and synchronous test sets.

\subsection{Source separation}
\label{sec:exp-separation}

Standard music source-separation models typically target vocals, drums, bass,
and other, collapsing instruments outside that taxonomy into a catch-all stem.
JazzSAMBA enables fine-tuning separators for jazz-combo stems like piano and
horns.
We fine-tune HT-Demucs~6s~\cite{rouard2022hybrid}, a
hybrid waveform-spectrogram Transformer with a dedicated piano stem;
Open-Unmix~\cite{stoter2019openunmix}, a spectrogram-masking LSTM baseline;
and BS-RoFormer~SW~\cite{lu2024bsroformer}, a band-split RoPE Transformer, on
asynchronous JazzSAMBA takes to separate stem-sum mixtures of drums, bass,
piano, and horns.
Horns are initialized from the pretrained ``other'' stem; Open-Unmix has no
piano stem, so we initialize its piano head from a copy of pretrained
``other''.
We compare zero-shot against fine-tuning on JazzSAMBA or
ChoraleBricks~\cite{balke2025choralebricks}, which provides isolated wind
parts including flute, oboe, clarinet, trumpet, saxophone, baritone,
trombone, and tuba, to test whether wind stems transfer.
Training uses shared mix-and-separate $4$\,s crops: we RMS-normalize each
stem, apply a random per-stem gain, sum to a mixture, then peak-normalize the
mixture and matched targets together.
We report full-track SI-SDR~\cite{leroux2019sdr}, which
measures how closely an estimate matches a reference while ignoring overall loudness,
on the asynchronous test set
($n{=}5$) with early stopping on validation loss
(Table~\ref{tab:separation}).

\begin{table}[t]
  \centering
  \caption{Asynchronous-test SI-SDR~$\uparrow$ (dB), full-track ($n{=}5$).
  Dashes mark unfair zero-shot or ChoraleBricks cells (no matching stem
  supervision). We bold the best reportable score within each model.}
  \label{tab:separation}
  \footnotesize
  \setlength{\tabcolsep}{3.0pt}
  \begin{tabular}{@{}ll|cccc@{}}
    \toprule
    \textbf{Model} & \textbf{FT} & \emph{Drums} & \emph{Bass} & \emph{Piano} & \emph{Horns} \\
    \midrule
    \multirow{3}{*}{HT-Demucs~6s}
      & --- & 13.01 & 15.07 & 4.92 & --- \\
      & JazzSAMBA & \textbf{15.09} & \textbf{17.03} & \textbf{9.42} & \textbf{14.62} \\
      & ChoraleBricks & --- & 15.02 & --- & 8.98 \\
    \midrule
    \multirow{3}{*}{Open-Unmix}
      & --- & 5.24 & 5.63 & --- & --- \\
      & JazzSAMBA & \textbf{8.37} & \textbf{9.74} & \textbf{6.07} & \textbf{9.38} \\
      & ChoraleBricks & --- & 5.63 & --- & $-$0.66 \\
    \midrule
    \multirow{3}{*}{BS-RoFormer~SW}
      & --- & \textbf{19.12} & \textbf{20.83} & \textbf{18.31} & \textbf{21.57} \\
      & JazzSAMBA & 18.97 & 20.62 & 17.54 & 21.15 \\
      & ChoraleBricks & --- & 17.59 & --- & 2.00 \\
    \bottomrule
  \end{tabular}
\end{table}

Zero-shot BS-RoFormer is strongest on reportable stems; JazzSAMBA fine-tuning
keeps it within about $0.8$\,dB, so zero-shot remains the headline BS result.
JazzSAMBA fine-tuning makes horns reportable for HT-Demucs and piano and
horns reportable for Open-Unmix, and it also improves drums and bass on those
models.
ChoraleBricks fine-tuning does not improve JazzSAMBA test scores (e.g., BS
horns $21.57{\to}2.00$\,dB), suggesting limited transfer from wind chorales
to this jazz-combo setting.

\subsection{Chart-conditioned accompaniment}
\label{sec:exp-live}

We demonstrate how researchers may benchmark live music accompaniment using
JazzSAMBA by comparing chart conditioning against listen-only baselines.
We fine-tune StreamGen~\cite{wu2025streaming}, a streaming latent model for
stem-held-out accompaniment, with a causal decoder ($t_f{=}0$) so each frame
depends only on past mix audio and optional chart features, suited to live
accompaniment.
A four-way ablation over listen-only, $+$sections, $+$chords, and $+$both
trains specialists on both protocols with $10$\,s crops and optional
oracle-timed JazzSAMBA charts.
At inference we free-run $60$\,s from up to three form landmarks per take
(head-in and the first two soloist entries), conditioning only on the
non-target mix and chart.
Table~\ref{tab:accompaniment} reports COCOLA~\cite{ciranni2025cocola} and
Beat Alignment F1~\cite{pasini2026liveband,foscarin2024beatthis} as
mean over landmarks, for asynchronous ($n{=}5$) and synchronous
($n{=}3$) test songs.
COCOLA scores contrastive coherence between the non-target condition mix and
the generated stem; Beat Alignment F1 compares an oracle bar-grid metronome
from the take annotations to beats detected on the generated stem.

Across both protocols, chart masks change Beat F1 more than COCOLA: Beat F1
swings farther across conditions and the best mask often changes by stem,
while $+$chords often edges StreamGen's COCOLA over listen-only, showing that
JazzSAMBA's annotations can discriminate conditioning strategies.
Synchronous Beat F1 is somewhat lower under live bleed, while COCOLA remains
comparable to the asynchronous setting.
Live audio accompaniment remains a difficult task; these results are one
baseline where JazzSAMBA's stems and annotations help.

\begin{table}[t]
  \centering
  \caption{Accompaniment scores on $60$\,s free-run from form landmarks
  (asynchronous test $n{=}5$, synchronous test $n{=}3$; mean over landmarks).
  D, B, P, and H denote drums, bass, piano, and horns.
  Ground Truth scores the annotated target stem against the condition mix.
  We bold the best score per column among chart conditions
  (excluding Ground Truth).}
  \label{tab:accompaniment}
  \footnotesize
  \setlength{\tabcolsep}{2.0pt}
  \begin{tabular}{@{}ll|cccc|cccc@{}}
    \toprule
    & & \multicolumn{4}{c|}{\textbf{COCOLA}~$\uparrow$}
      & \multicolumn{4}{c}{\textbf{Beat F1}~$\uparrow$} \\
    \cmidrule(lr){3-6} \cmidrule(l){7-10}
    \textbf{Protocol} & \textbf{Condition} & \emph{D} & \emph{B} & \emph{P} & \emph{H} & \emph{D} & \emph{B} & \emph{P} & \emph{H} \\
    \midrule
    \multirow{5}{*}{Asynchronous}
      & \emph{Ground Truth}
        & 55.6 & 62.1 & 63.4 & 64.5
        & 1.00 & 0.80 & 0.48 & 0.29 \\
      & Listen-Only
        & 52.2 & 60.4 & 59.8 & 61.4
        & 0.23 & \textbf{0.16} & 0.13 & 0.02 \\
      & $+$Sections
        & 50.7 & 59.8 & 59.7 & 61.9
        & 0.29 & 0.09 & 0.13 & 0.05 \\
      & $+$Chords
        & \textbf{52.5} & \textbf{60.6} & \textbf{60.0} & \textbf{62.3}
        & 0.30 & 0.09 & \textbf{0.15} & \textbf{0.06} \\
      & $+$Both
        & 51.4 & 60.4 & 59.9 & 61.7
        & \textbf{0.34} & 0.10 & 0.12 & 0.04 \\
    \midrule
    \multirow{5}{*}{Synchronous}
      & \emph{Ground Truth}
        & 58.4 & 61.2 & 62.9 & 65.9
        & 0.97 & 0.74 & 0.67 & 0.78 \\
      & Listen-Only
        & 53.2 & \textbf{60.3} & 58.6 & 60.8
        & 0.26 & \textbf{0.12} & 0.09 & 0.05 \\
      & $+$Sections
        & 52.8 & 58.9 & 58.3 & \textbf{61.4}
        & 0.27 & 0.06 & 0.14 & 0.01 \\
      & $+$Chords
        & \textbf{53.5} & 59.1 & \textbf{58.9} & 60.9
        & 0.27 & 0.09 & 0.09 & \textbf{0.05} \\
      & $+$Both
        & 51.3 & 59.9 & 58.7 & 60.9
        & \textbf{0.31} & 0.07 & \textbf{0.15} & 0.00 \\
    \bottomrule
  \end{tabular}
\end{table}

\section{Conclusion}
\label{sec:conclusion}

JazzSAMBA is the first jazz-standard dataset with per-player stems and
annotations for chords, sections, and soloists.
It includes 76 standards with multi-take audio and MIDI in asynchronous and
synchronous settings, performed by eight musicians.
Paired with performance-aligned charts, it supports music information retrieval
and generation.
We show that it enables jazz-combo source separation for piano and horns, and
benchmarks live chart-conditioned accompaniment.
We release JazzSAMBA hoping others will push beyond these baselines toward
interactive chart-following accompaniment, form-aware combo analysis, and
models that handle take-to-take and live-ensemble variation.

\section{Compliance with Ethical Standards}
Performing musicians are credited as coauthors and consented to the recording
and public release of the dataset.
Recordings are original studio performances.
The dataset contains those recordings and our annotations only.

\bibliographystyle{IEEEbib}
\makeatletter
\let\jazzsamba@thebibliography\thebibliography
\renewcommand{\thebibliography}[1]{%
  \jazzsamba@thebibliography{#1}%
  \setlength{\itemsep}{0pt plus 0.05pt}%
  \setlength{\parsep}{0pt}%
}
\makeatother
\bibliography{main}

\end{document}